\documentclass[useAMS,usenatbib]{mn2e}
\usepackage{epsfig}
\usepackage{amsmath}

\newcommand{\be}{\begin{equation}}
\newcommand{\beq}{\begin{equation}}
\newcommand{\ba}{\begin{eqnarray}}
\newcommand{\ee}{\end{equation}}
\newcommand{\eeq}{\end{equation}}
\newcommand{\ea}{\end{eqnarray}}

\def\lsim{~\rlap{$<$}{\lower 1.0ex\hbox{$\sim$}}}

\def\gsim{~\rlap{$>$}{\lower 1.0ex\hbox{$\sim$}}}

\title[The Histories of Reionization in Hierarchical Galaxy Formation Models]{Variation in the escape fraction of ionising photons from galaxies and the redshifted 21-cm power spectrum during reionization}
\author[Han-Seek Kim et al.]
       {Han-Seek~Kim$^{1}$\thanks{hansikk@unimelb.edu.au}, J. Stuart B.~Wyithe$^{1,3}$, Jaehong~Park$^{1}$, C. G.~Lacey$^{2}$\\
       $^1$School of Physics, The University of Melbourne, Parkville, VIC 3010, Australia\\
      $^2$Institute for Computational Cosmology, Department of Physics, University of Durham, South Road, Durham DH1 3LE, UK\\
  	$^3$ARC Centre of Excellence for All-sky Astrophysics (CAASTRO)}
\date{}

\pagerange{\pageref{firstpage}--\pageref{lastpage}}
\pubyear{2013} 

\begin{document}

\maketitle

\label{firstpage}

\begin{abstract}

\noindent 

The observed power spectrum of redshifted 21cm fluctuations is known to be sensitive to the astrophysical properties of the galaxies that drove reionization. Thus, detailed measurements of the 21cm power spectrum and its evolution could lead to measurements of the properties of early galaxies that are otherwise inaccessible. In this paper, we study the effect of mass and redshift dependent escape fractions of ionizing radiation on the ability of forthcoming experiments to constrain galaxy formation via the redshifted 21cm power spectrum. We use a model for reionization which combines the hierarchical galaxy formation model GALFORM  implemented within the Millennium-II dark matter simulation, with a semi-numerical scheme to describe the resulting ionization structure. 
Using this model we show that the structure and distribution of ionised regions at fixed neutral fraction, and hence the slope and amplitude of the 21 cm power spectrum, is dependent on the variation of ionising photon escape fraction with galaxy mass and redshift. However, we find that the influence of the unknown escape fraction and its evolution is smaller than the dominant astrophysical effect provided by SNe feedback strength in high redshift galaxies. The unknown escape fraction of ionizing radiation from galaxies is therefore unlikely to prevent measurement of the properties of high redshift star formation using observations of the 21cm power spectrum.

\end{abstract}

\begin{keywords}
Cosmology: theory; diffuse radiation; dark ages, reionization, first stars; Galaxies: high-redshift
\end{keywords}

\section{Introduction}
 
In recent years a great deal of
theoretical attention has focused on modeling the effect of galaxies on
the reionization of the IGM. In large modern simulations, the most common approach
 is to begin with an N-body simulation to generate a distribution of dark matter
halos \citep[e.g.][]{ciardi2003,sokasian2003,iliev2007,zahn2007,trac2007,shin2008,Il08,trac2008}. A prescription is then used to relate dark matter halo mass to ionizing luminosity. Following this step, radiative transfer methods (most commonly ray-tracing algorithms) are employed to model the generation of ionized structure on large scales. The radiative transfer is normally run with lower resolution than the N-body code for computational
efficiency.

An important outcome from the large cosmological volumes attained by 
modern numerical simulations has been the prediction of 21~cm
signals that will be observable using forthcoming low frequency arrays
\cite[e.g.][]{mellema2006,lidz2008}. \citet{lidz2008} argue that 
first generation low frequency radio telescopes like the Murchison Widefield Array\footnote{http://www.haystack.mit.edu/ast/arrays/mwa/} and the Low Frequency Array\footnote{http://www.lofar.org/} will have sufficient sensitivity to measure the
redshift evolution in the slope and amplitude of the 21~cm power
spectrum. One of the main limitations in modelling of reionization is the physics of the ionizing sources. Most studies have used very simple prescriptions to assign ionizing luminosities to dark matter halos. It has been shown that it is then possible to constrain the parameters for these simple prescriptions \citep[e.g.][]{barkana2008}. However, an important open question is the degree to which the important astrophysics governing formation and evolution of high redshift galaxies is accessible via observations of the 21cm power spectrum. 

Several studies have previously addressed the issue of realistic modelling of high redshift galaxies and their role in reionization. For example, \citet[][]{Theuns2011} \citep[see also ][]{Benson2006,Lacey2011} used the hierarchical galaxy formation code, GALFORM \citep{Cole2000,Baugh2005,Bower2006}, implemented within Monte-Carlo merger trees to evaluate the ionizing photon budget. They found that although galaxies should produce sufficient ionizing photons to complete reionization, most of the galaxies responsible would be below the detection threshold of current surveys. However, these studies were restricted to global evolution of the ionized fraction, and did not address the ionization structure of the IGM. Most recently, \citet[][]{Kim2012a} have combined the GALFORM model implemented within the Millennium-II dark matter simulation, with a semi-numerical scheme to describe the resulting ionization structure.
In this paper we extend this model to consider an escape fraction for ionising photons that depends on redshift and host halo mass. 

The escape fraction of ionizing photons from their host galaxies is one of the most important unknowns for the reionization history.
Observational estimates show a broad range of escape fraction values from a few percent in the local Unvierse \citep{hurwitz1997,bland1999,putman2003}, to a few percent or a few tens of percent at redshift $z\sim$ 1-3 \citep{inoue2006,shapley2006,chen2007,siana2007}. However, there are no observational constraints on the escape fraction during the epoch of reionization. Theoretically, the escape fraction is predicted to span a very broad range from 0.01 to 1 \citep{wood2000,ricotti2000,Benson2002,fujita2003,sokasian2003,ciardi2003,WL07,WC09,Theuns2011,kuhlen2012}. 
 
In this paper we incorporate a variable escape fraction for ionising photons within the model of \citet[][]{Kim2012a}, and predict the  redshifted 21cm power spectrum for the resulting reionization histories. We begin in \S~\ref{modell} and \S~\ref{scheme} by describing the implementation of GALFORM, and our method for modelling the ionization structure. Then, in \S~\ref{REhistories} we present variant models which lead to different reionization histories, and present a discussion of the dependence of the 21cm power spectrum on the escape fraction models in \S~\ref{PS}.  We finish with some conclusions in \S~\ref{Summary}.

\section{The model} \label{modell}
In this section we summarise the theoretical galaxy formation modelling based on \citep{Kim2012a} that is used in our analysis in order to describe the new features for this paper.

\subsection{The GALFORM galaxy formation model}\label{GFM}

We compute the formation and evolution of galaxy properties using the hierarchical galaxy formation model GALFORM. 
GALFORM includes a range of processes that are thought to be important for galaxy formation from dark matter merger histories to baryonic physics such as cooling, star formation, feedback processes, and chemical enrichment. A comprehensive overview of GALFORM can be found in \citet[][]{Cole2000}, with an updated discussion in the review by \citet[][]{Baugh2006}. 
Feedback processes (including AGN, SNe and photo-ionization) during galaxy formation play very important roles in shaping the luminosity functions predicted by GALFORM \citep[][]{Cole2000,Benson2002,Baugh2005,Bower2006,Kim2011,Kim2012b}.  
In this paper, we use the version of GALFORM described in \citet{Lagos2012}. The Millennium-II simulation on which the model is based has a cosmology with fractional mass and dark energy densities of $\Omega_{\rm m}=0.25$ , $\Omega_{\rm b}=0.045$ and $\Omega_{\Lambda}$=0.75, a dimensionless Hubble constant of $h$=0.73, and a power spectrum normalisation of $\sigma_{8}$=0.9. The particle mass in the simulation is 6.89$\times$10$^{6}$h$^{-1}{\rm M_{\odot}}$ and we detect halos down to 20 particles in the simulation box of side length $L=100h^{-1}$Mpc.

\section{Semi-Numerical sheme to calculate the evolution of the ionised structure}
\label{scheme}

\citet{MF07} introduced an approximate but efficient method for
simulating the reionization process.  In this paper we apply a semi-numerical technique to find the ionization structure resulting from GALFORM galaxies within the Millennium-II dark matter simulation, following the procedure described in \citet{Kim2012a}.

We assume the number of photons produced by galaxies within a small volume (or cell) of the simulation, and which enter the IGM to participate in reionization to be
\begin{equation}\label{nphotons}
N_{\rm \gamma, cell}={\it f}_{\rm esc}\int^{t_{z}}_{0}\dot{N}_{\rm Lyc,cell}(t)~dt,
\end{equation}
where $f_{\rm esc}$ is the escape fraction of ionizing photons produced by stars in a galaxy. The total Lyman continuum luminosity of the $N_{\rm cell}$ galaxies within the cell expressed as the emission rate of ionizing photons { (i.e.  units of photons/s)} is
\begin{equation}
\dot{N}_{\rm Lyc,cell}(t) = \sum_{i=1}^{N_{\rm cell}} \dot{N}_{\rm Lyc, \it i}(t),
\end{equation}
where
\begin{equation}
\dot{N}_{\rm Lyc, \it i}(t)=\int^{\infty}_{\nu_{\rm thresh}}{L_{\nu,i}(t) \over h\nu} {\rm d\nu},
\end{equation}
{\it L$_{\nu,i}$} is the spectral energy distribution of galaxy $i$, and $\nu_{\rm thresh}$ is the Lyman-limit frequency ($h\nu_{\rm thresh}$ = 13.6 eV).
Note that the number of photons produced per baryon in long-lived stars and stellar remnants depends on the IMF and metalicity ({\it Z}).
We assume the total Lyman continuum luminosity in a cell at redshift $z_{i}$ to be constant until the next snapshot at redshift $z_{i+1}$, and calculate the number of photons produced in the cell between $z_{i}$ and $z_{i+1}$ 
as $\dot{N}_{\rm Lyc,cell}(t_{z_{i}})  \times (t_{z_{i+1}}-t_{z_{i}})$.

We then calculate the ionization fraction within each cell according to  
\begin{equation}
\label{Qvalue}
Q_{\rm cell}=\left[{N_{\rm \gamma, cell} \over (1+F_{\rm c})N_{\rm HI, cell}}\right],
\end{equation}
where $F_{\rm c}$ denotes the mean number of recombinations per hydrogen atom up to reionization and $N_{\rm HI, cell}$ is the number of neutral hydrogen atoms within a cell. The latter quantity is calculated as 
\begin{equation}
\label{nHI}
N_{\rm HI, cell}={\bar{n}_{\rm HI}(\delta_{\rm DM,cell}+1)V_{\rm cell}},
\end{equation}
where we assume that the overdensity of neutral hydrogen follows the dark matter (computed based on the Millennium-II simulation density field), $\bar{n}_{\rm HI}$ is the mean comoving number density of hydrogen atoms, and $V_{\rm cell}$ is the comoving volume of the cell. Self-reionization of a cell occurs when $Q_{\rm cell}>1$. We divide the Millennium-II simulation box into 256$^{3}$ cells, yielding cell side lengths of 0.3906$h^{-1}$Mpc and comoving volumes of 0.0596$h^{-3}$Mpc$^{3}$.  

Theoretical prediction of the parameters $F_{\rm c}$ and $f_{\rm esc}$ in Equations~(\ref{nphotons}) and~(\ref{Qvalue}) from first principles is complicated, and their values are not known. The recombination parameter $F_{\rm c}$ is related to the density of the IGM on small scales, while $f_{\rm esc}$ depends on the details of the high redshift ISM. Previous work using GALFORM suggested the value $(1+F_{\rm c})/f_{\rm esc}\sim10$ \cite[][]{Benson2001,Theuns2011,Kim2012a} to fit observational constraints on reionization. Here, we vary the value of $f_{\rm esc}$, assuming $F_{\rm c}$=0.5, as a function of redshift or host dark matter halo mass to see the effect of different histories of reionization on the 21cm power spectrum.

Based on equation~(\ref{Qvalue}), individual cells can have $Q_{\rm cell}>1$. On the other hand, cells with $Q_{\rm cell}<1$ may be ionized by photons produced in a neighbouring cell. In order to find the extent of ionized regions we therefore filter the $Q_{\rm cell}$ field using a sequence of real space top hat filters of radius $R$ (with $0.3906<R<100h^{-1}$Mpc), producing one smoothed ionization field $Q_R$ per radius. At each point in the simulation box we find the largest $R$ for which the filtered ionization field is greater than unity (i.e. ionized with $Q_R>1$). All points within the radius $R$ around this point are considered ionized. This procedure forms the position dependent ionization fraction $0\leq Q\leq1$, which describes the ionization structure of the IGM during reionization.
 
In this paper we restrict our attention to analyses that assume the spin temperature of hydrogen is coupled to the kinetic temperature of an IGM that has been heated well above the CMB temperature (i.e. $T_{\rm s}\gg T_{\rm CMB}$). This condition should hold during the later stages of the reionization era \citep[$z\la9$, ][]{S+07}. In this regime there is a proportionality between the ionization fraction and 21~cm intensity, and the 21~cm brightness temperature contrast may be written as
\begin{equation}
\label{Tb}
\Delta T(z)=T_{0}(z)\left[1-Q\right]\left(1+\delta_{\rm DM,cell}\right),
\end{equation} 
where $T_{0}(z)=23.8\left(\frac{1+z}{10}\right)^\frac{1}{2}\mbox{mK}$. We have ignored the contribution to the amplitude from velocity gradients, and assumed as before that the hydrogen overdensity follows the dark matter ($\delta_{\rm DM,cell}$).

\section{Modelling the histories of reionization}\label{REhistories}
In this paper, we assume several different models for $f_{\rm esc}$ as defined in \S~\ref{scheme}. Specifically we model dependencies of the escape fraction with halo mass and redshift using

\begin{equation}\label{ModelEq}
f_{\rm esc}=A \times { \left(1+z \over 7\right)}^{\alpha} \left({{\rm log_{10}(M_{halo}/h^{-1}M_{\odot})/10}}\right)^{\beta},
\end{equation} 
where $A$ is normalisation parameter, $\alpha$ describes the redshift dependence, and $\beta$ describes the halo mass dependence. We force to $f_{\rm esc}$=0 when the value of Eq.~\ref{ModelEq} is negative and $f_{\rm esc}$=1 when the value greater than unity. We also consider a fiducial model with $f_{\rm esc}$=constant. We assume $F_{c}$ to equal 0.5 \citep{BL01,furlanetto2006}. To facilitate comparison, we assume that $A$ is normalized so that the ionization fraction at z=7.272 is $\left<x_{i}\right>$=0.5 for all models. In Table.~\ref{Parameters}, we show the parameters describing the 5 models used in this paper. These models are discussed below.

\begin{table}
\caption{
The values of selected parameters which are different in the models. The columns are as follows: (1) the name of the model, (2) Normarlization value to match $<x_{i}>$=0.5 at z=7.272, (3) index $\alpha$ describing redshift dependence, (4) index $\beta$ describing halo mass dependence.}
\label{Parameters}
\begin{tabular}{lccc}
\hline
\hline
 & $A$ & $\alpha$ & $\beta$ \\
\hline
Model 0 &0.5348 & 0 & 0\\
Model I &0.1488& 5 & 0\\
Model II & 1.6791 & -5 & 0\\
Model III& 0.3649& 0 & 5\\
Model IV & 0.8966 & 0 & -5 \\
\hline
\end{tabular}
\end{table}

\subsection{Redshift dependence modelling}
There is uncertainty in the redshift dependence of the escape fraction of ionising photons \citep{inoue2006,shapley2006,siana2010,kuhlen2012}.
Here, we model the effect of a redshift dependence of the escape fraction, $f_{\rm esc}$, on the evolution of the mass averaged ionization fraction using three different redshift dependences. 
\\

Model 0: $f_{\rm esc}$  is constant with redshift.

Model I: $f_{\rm esc}$  increases with redshift.

Model II: $f_{\rm esc}$  decreases with redshift.\\

\begin{figure*}
\includegraphics[width=6.5cm]{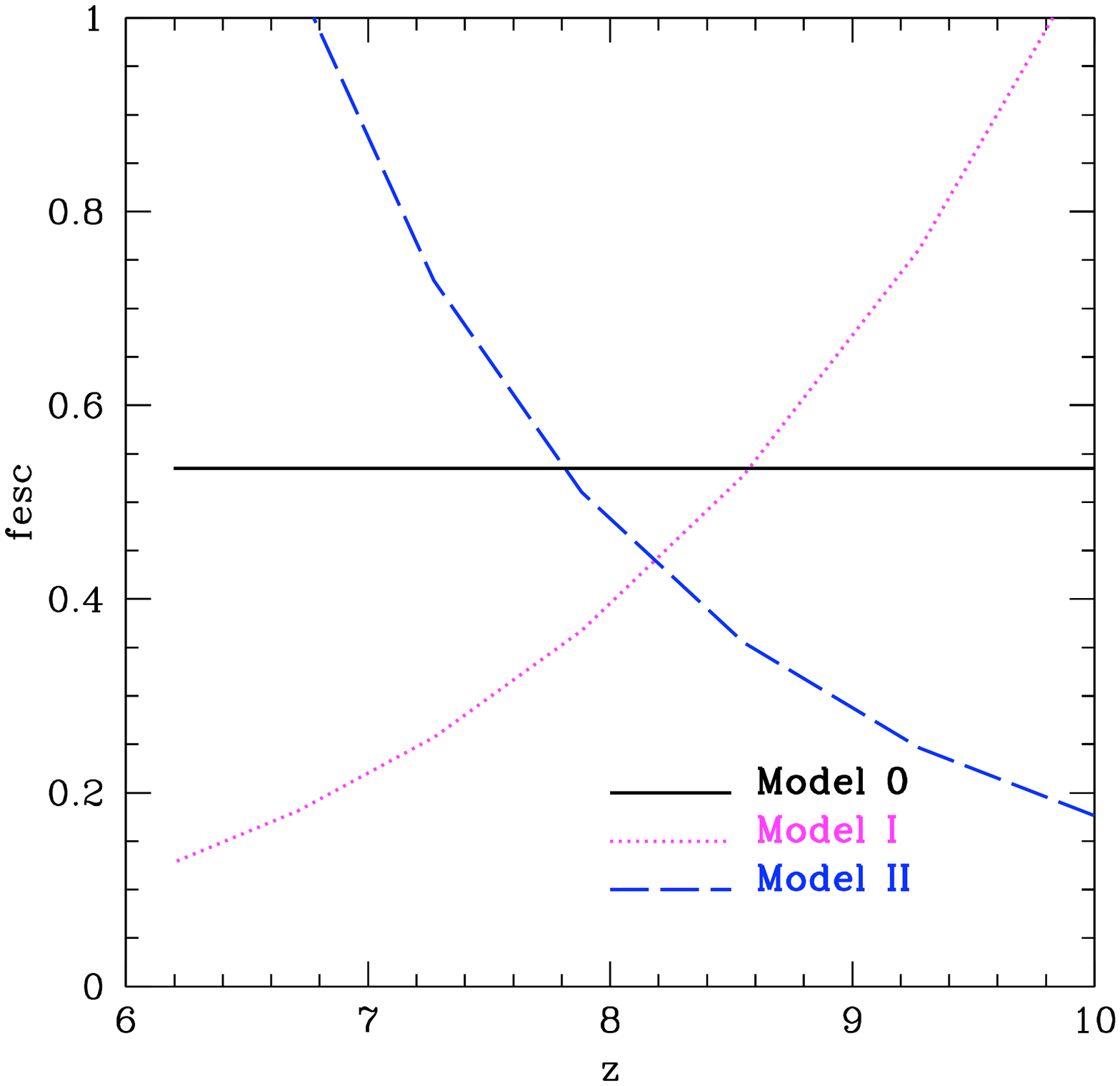}
\includegraphics[width=6.5cm]{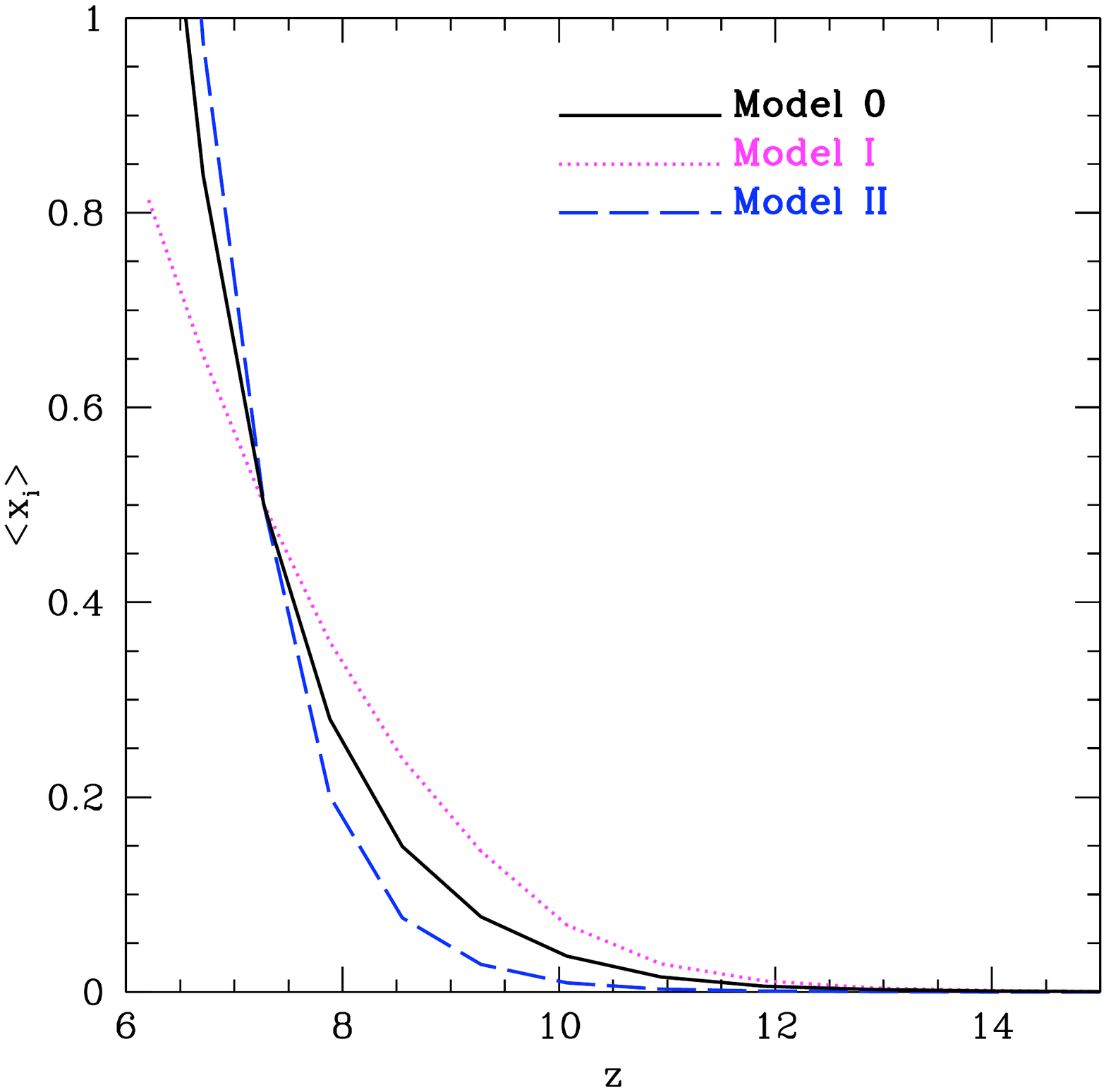}
\vspace{-1.5cm}
\caption{Left: The evolution of $f_{\rm esc}$ as a function of redshift for the three redshift-dependent models 0, I and II. Right:  The evolution of the mass averaged ionization fraction evolution for the models 0, I and II.} \label{Zdepmodels}
\end{figure*}
Figure~\ref{Zdepmodels} shows the assumed redshift evolution of $f_{\rm esc}$ (left panel), together with the resulting mass averaged ionization fraction ($\left<x_{i}\right>$) from the different models (right panel). Model~II completes reionization earlier than the other two models, and has the fastest evolution. On the other hand, Model~I completes reionization later than other two models and has the slowest evolution. 

\begin{figure*}
\includegraphics[width=6.5cm]{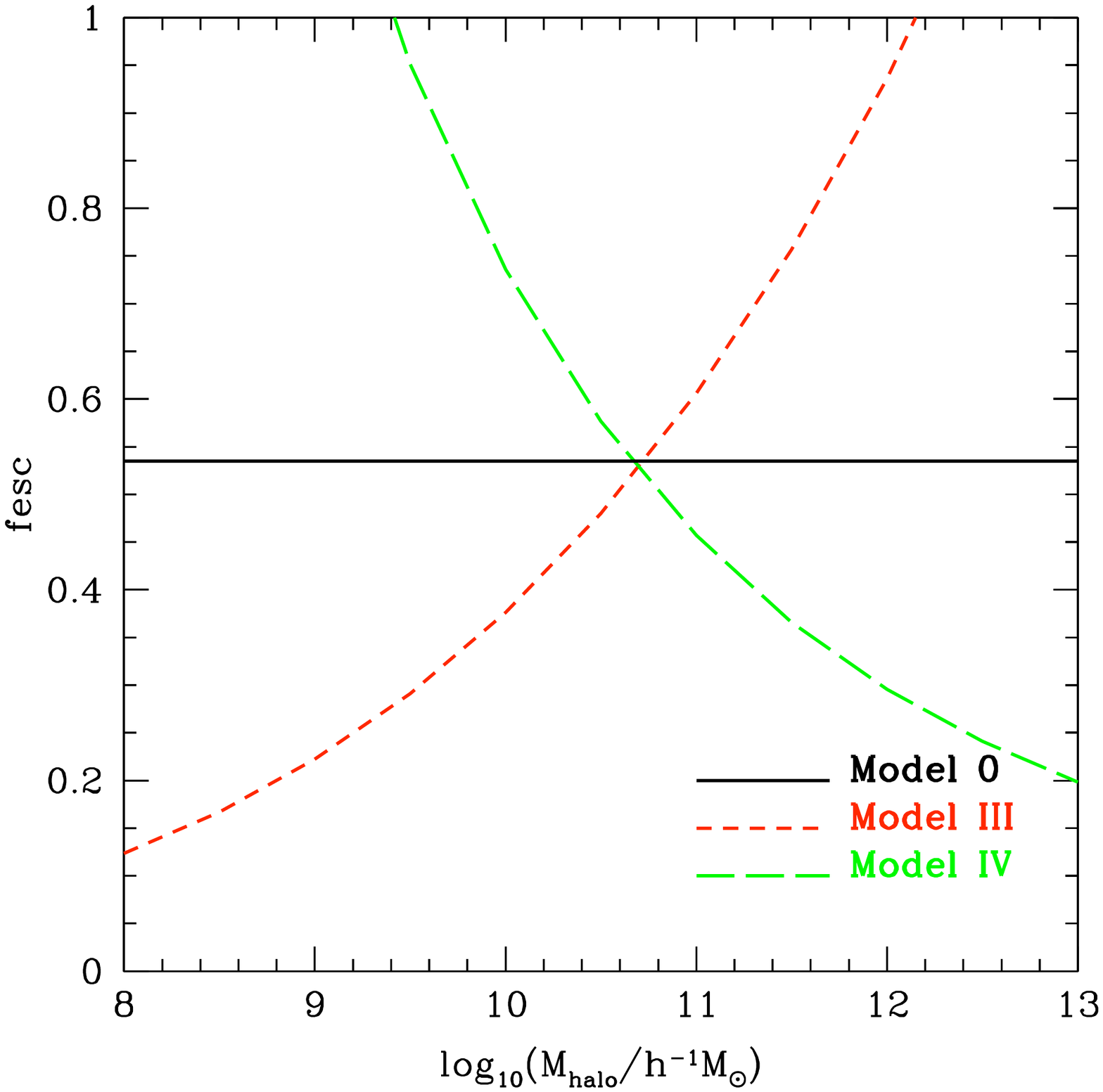}
\includegraphics[width=6.5cm]{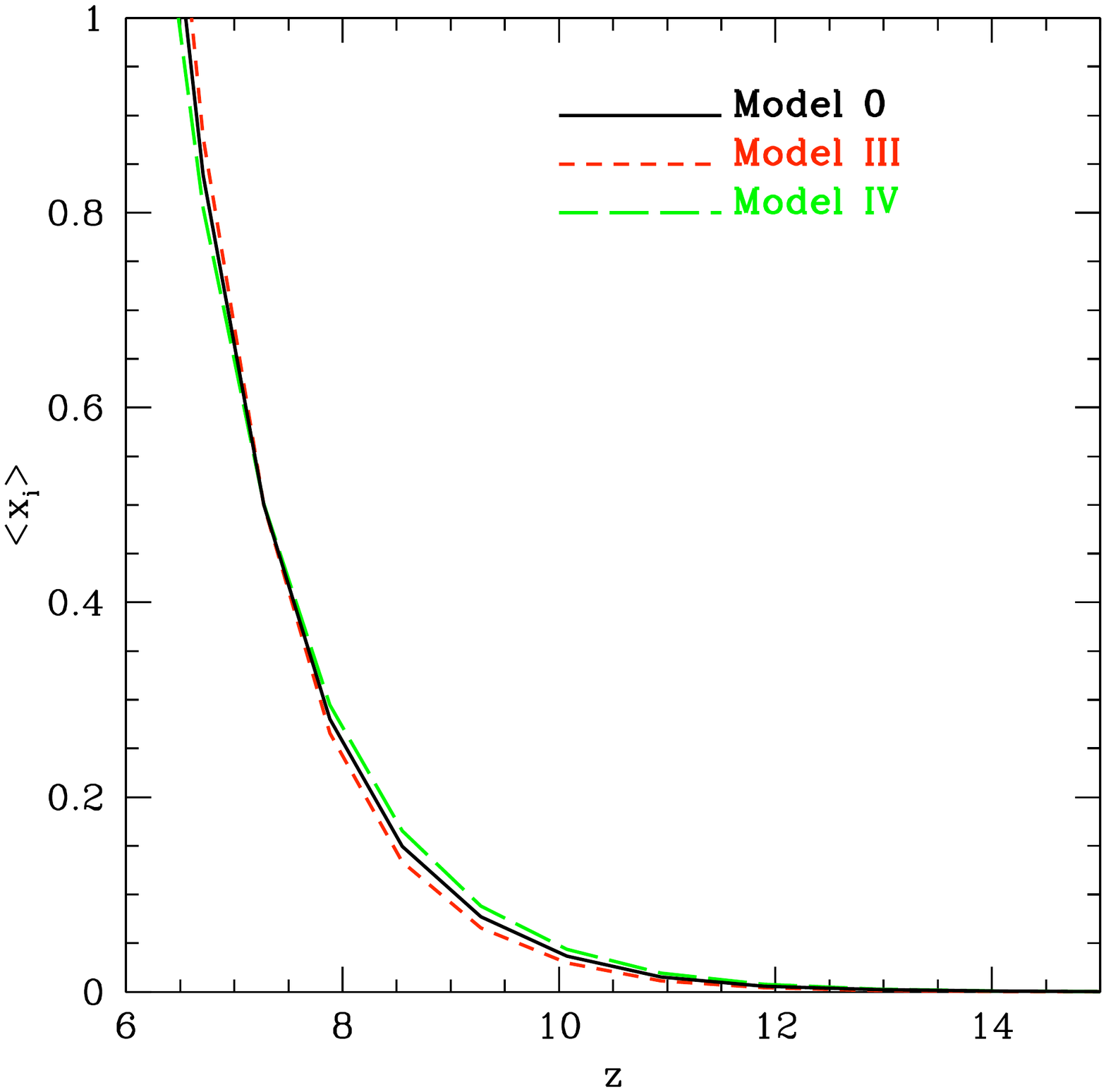}
\vspace{-1.5cm}
\caption{Left: The values of $f_{\rm esc}$ as a function of host dark matter halo mass for  two halo dependent models III and IV. Right: The evolution of the mass averaged ionization fraction for the two models III and IV.} \label{Mdepmodels}
\end{figure*}

\subsection{Halo mass dependence modelling} 
It is unknown whether the escape fraction of photons has a dependence on host halo mass \citep{gnedin2007,WC09,razoumov2010,yajima2011}.
Here we model the effect of a host halo mass dependence of escape fraction on the evolution of the mass averaged ionization fraction using two different assumptions for the host dark matter halo mass dependence. \\

Model III: $f_{\rm esc}$  increases with halo mass.

Model IV: $f_{\rm esc}$  decreases with halo mass.\\

Figure~\ref{Mdepmodels} shows the assumed dependence of $f_{\rm esc}$ on host dark matter halo mass (left panel) and the evolution of the resulting mass averaged ionization fractions ($\left<x_{i}\right>$). Models~III  and IV show similar evolution of ionization fraction. This is because the ionizations are dominated by galaxies near 10$^{10}{\rm h}^{-1}$M$_{\odot}$, above which galaxies are rare, and below which flux is suppressed by feedback processes. The difference between the ionization fraction evolution histories of the host halo mass dependent models is smaller than for the redshift dependent models in Figure.~\ref{Zdepmodels}.

\subsection{Modelling without supernovae feedback}
For comparison, we also include a model without supernovae (SNe) feedback that is otherwise the same as Model~0. This model is referred to as NOSN.  Note that we adjust $f_{\rm esc}$ for the NOSN model as a function of redshift so as to match the ionization fraction from Model~0. The resulting values of $f_{\rm esc}$=0.012, 0.017, 0.023, 0.032 and 0.041 correspond to redshifts of $z\sim$ 9.278, 8.550, 7.883, 7.272 and 6.712.

\section{The 21-cm power spectrum}
\label{PS}

The filtering procedure described in \S~\ref{scheme} provides  3-dimensional maps of the ionization structure and 21cm intensity within the Millennium-II Simulation box. From this cube we calculate the dimensionless 21-cm power spectrum 
\begin{equation}
\Delta^{2}(k)=k^3/(2\pi^2)P_{21}(k,z)/{T_{0}(z)}^{2}
\end{equation}
as a function of spatial frequency $k$, where $P_{21}(k)$ is the 3-dimensional power spectrum of 21cm brightness temperature $\Delta T(z)$   (described by eq.~(\ref{Tb})). 

\begin{figure*}
\begin{center}
\includegraphics[width=5.7cm]{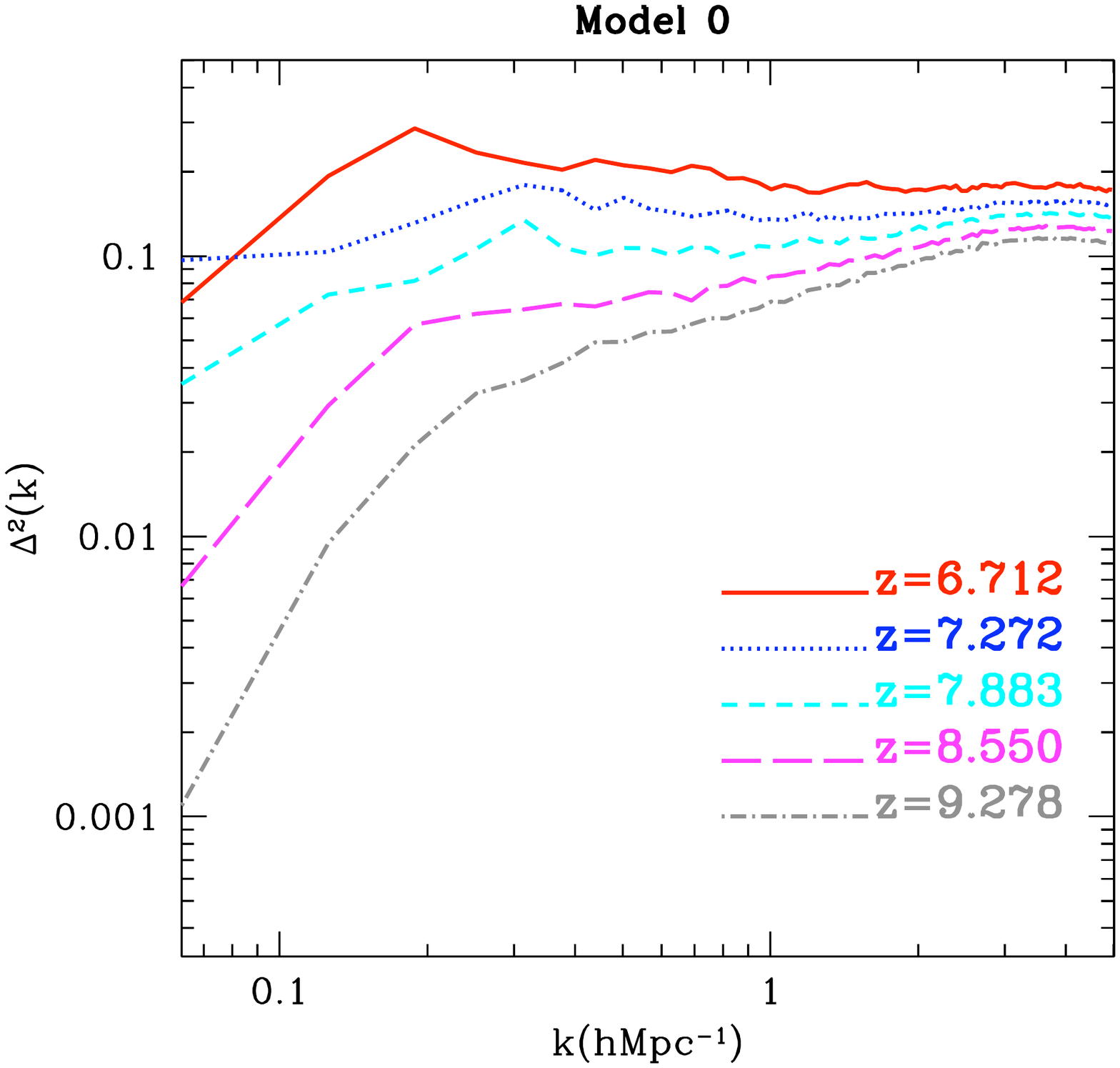}
\vspace{-2cm}
\includegraphics[width=5.7cm]{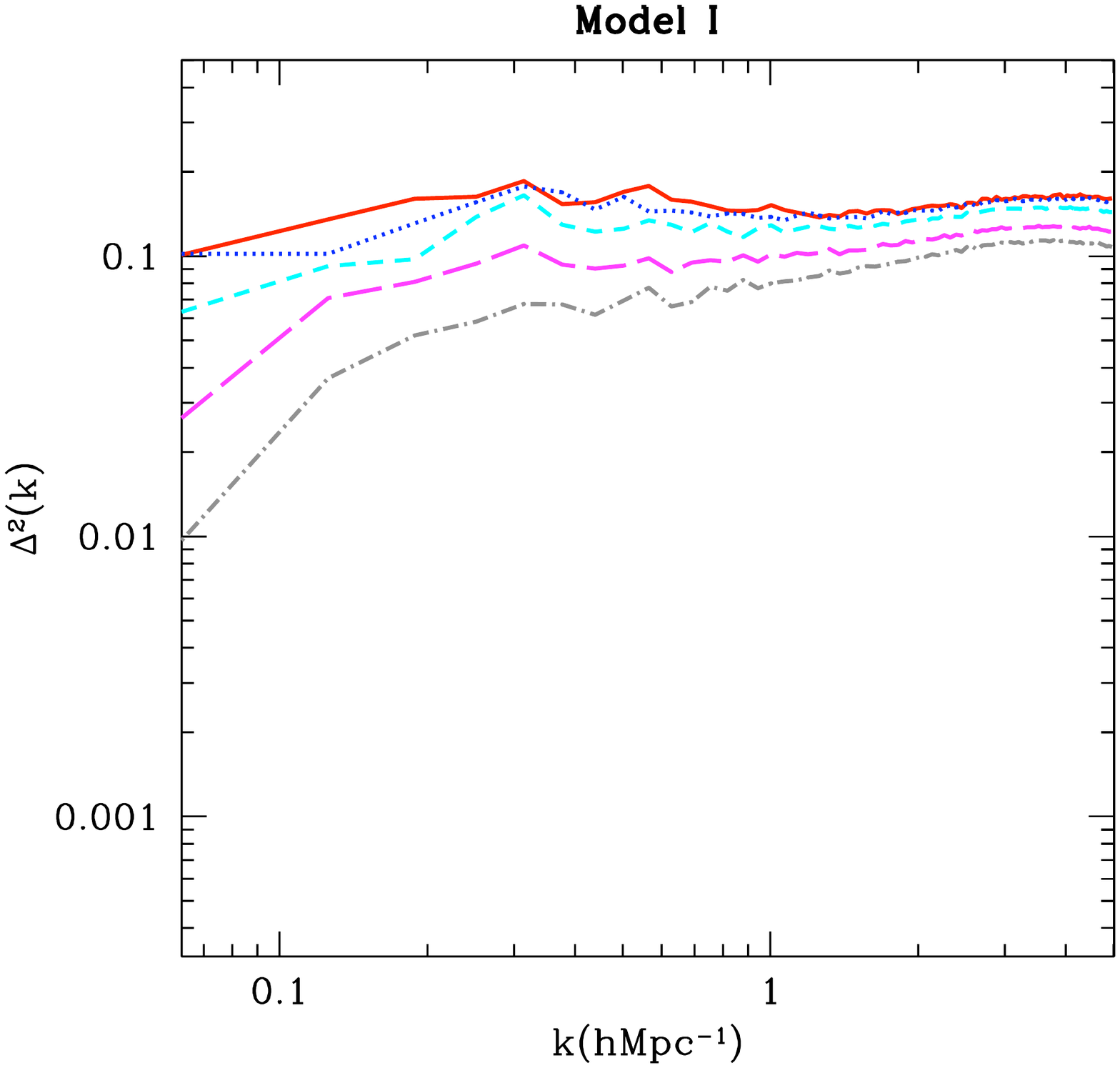}
\includegraphics[width=5.7cm]{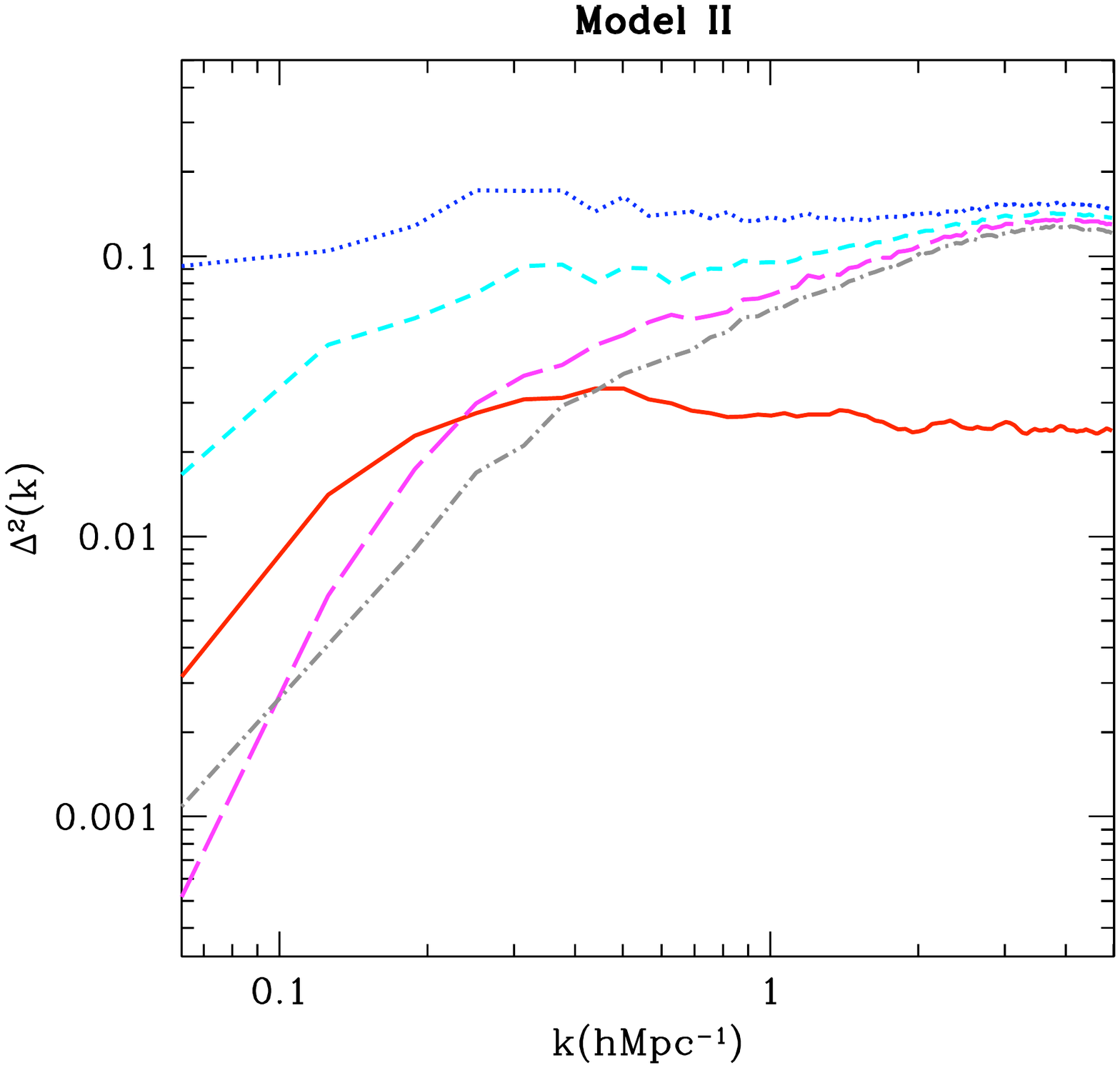}
\includegraphics[width=5.7cm]{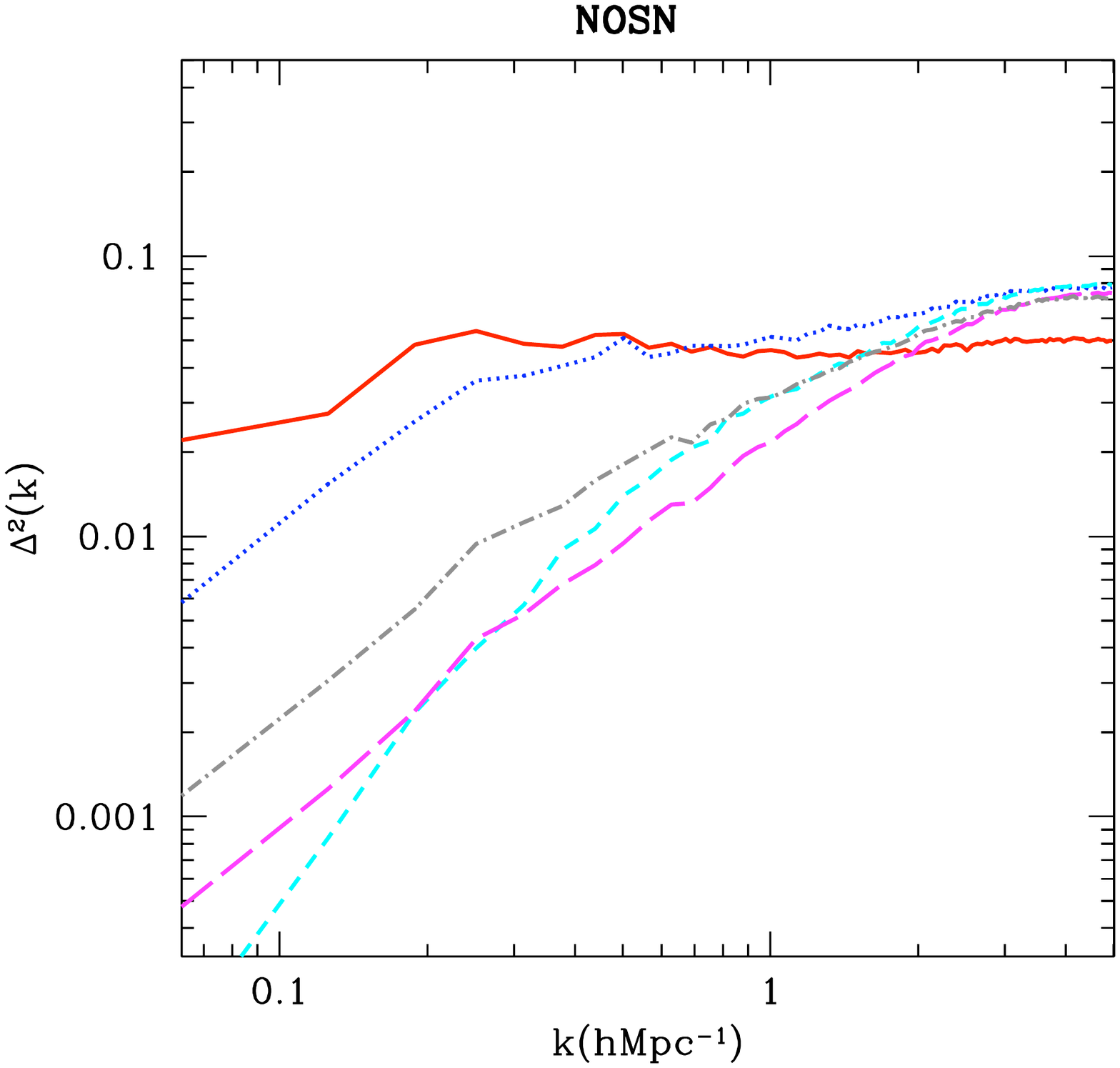}
\includegraphics[width=5.7cm]{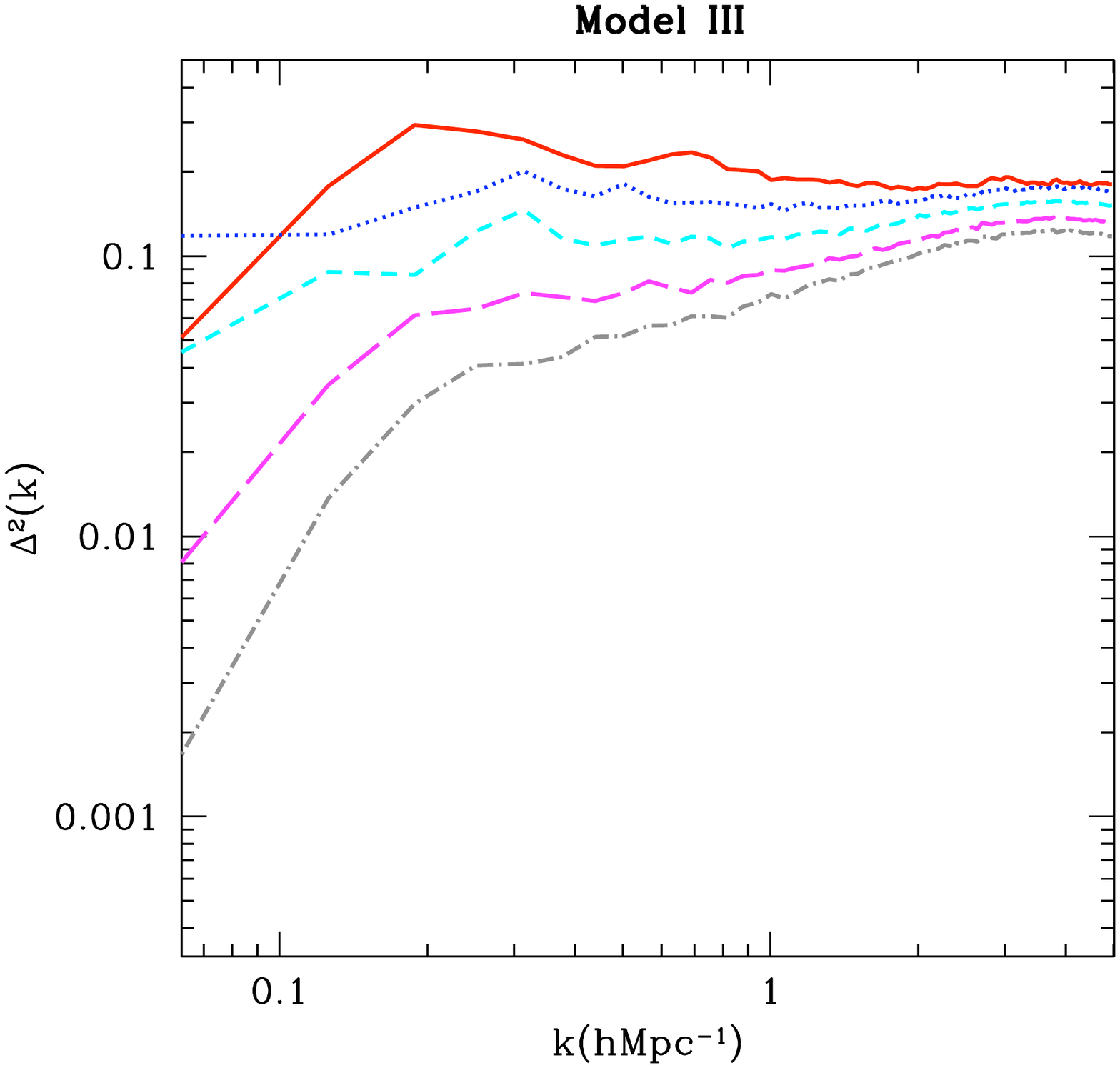}
\includegraphics[width=5.7cm]{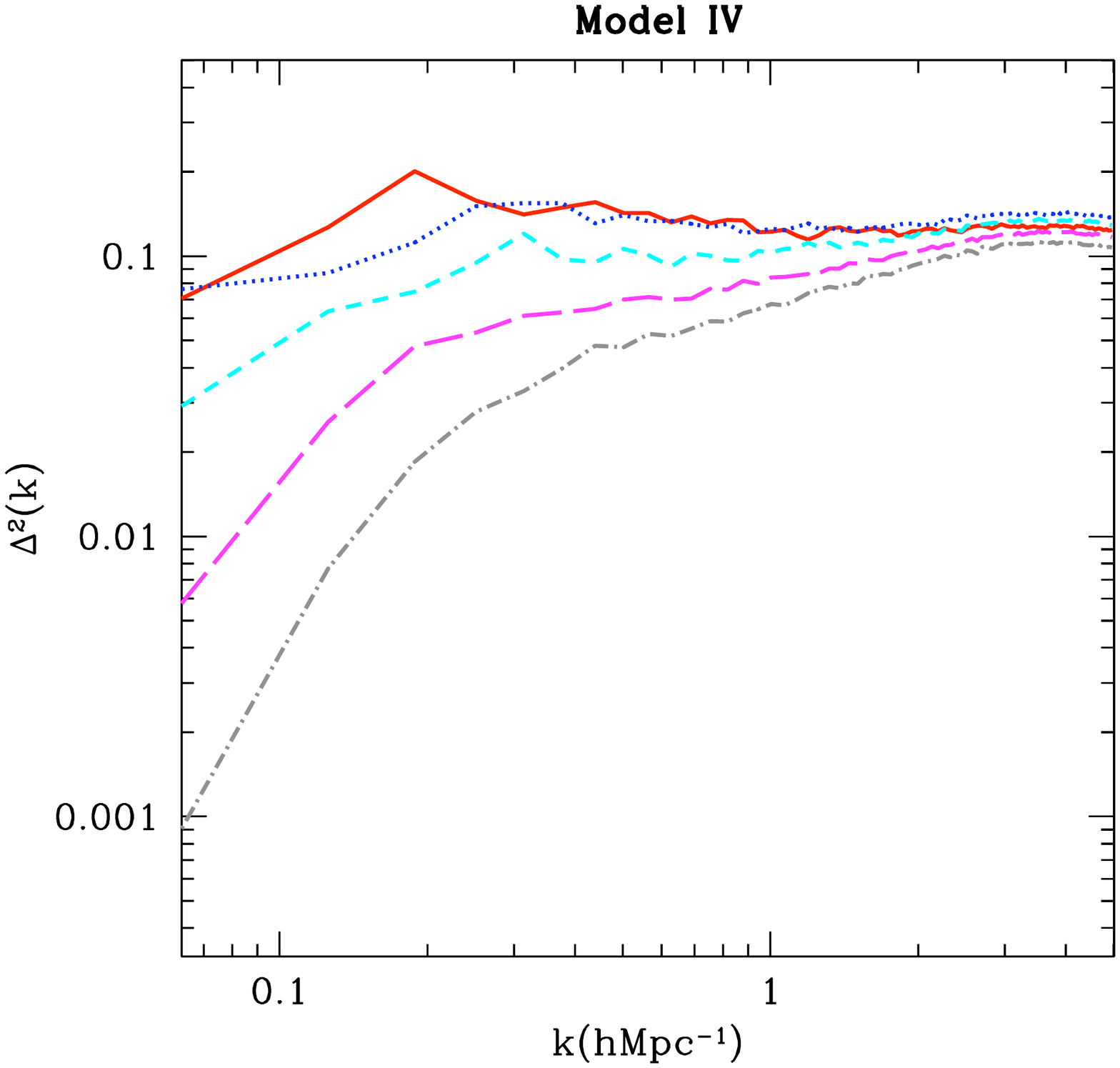}

\end{center}
\vspace{-2.cm}
\caption{The evolution of the predicted 21-cm dimensionless power spectra. Colours and line styles show different redshifts as shown by the key. 
} \label{PSA}
\end{figure*}

Figure~\ref{PSA} shows the evolution of the predicted dimensionless 21-cm power spectra for all models. The predicted power spectra show different amplitudes and shapes at the same redshift partly because each model has
reached a different stage of reionization, and partly because of the relative fraction of ionizing photons emitted from large verses small galaxies.  

\begin{figure*}
 \begin{center}
\includegraphics[width=8cm]{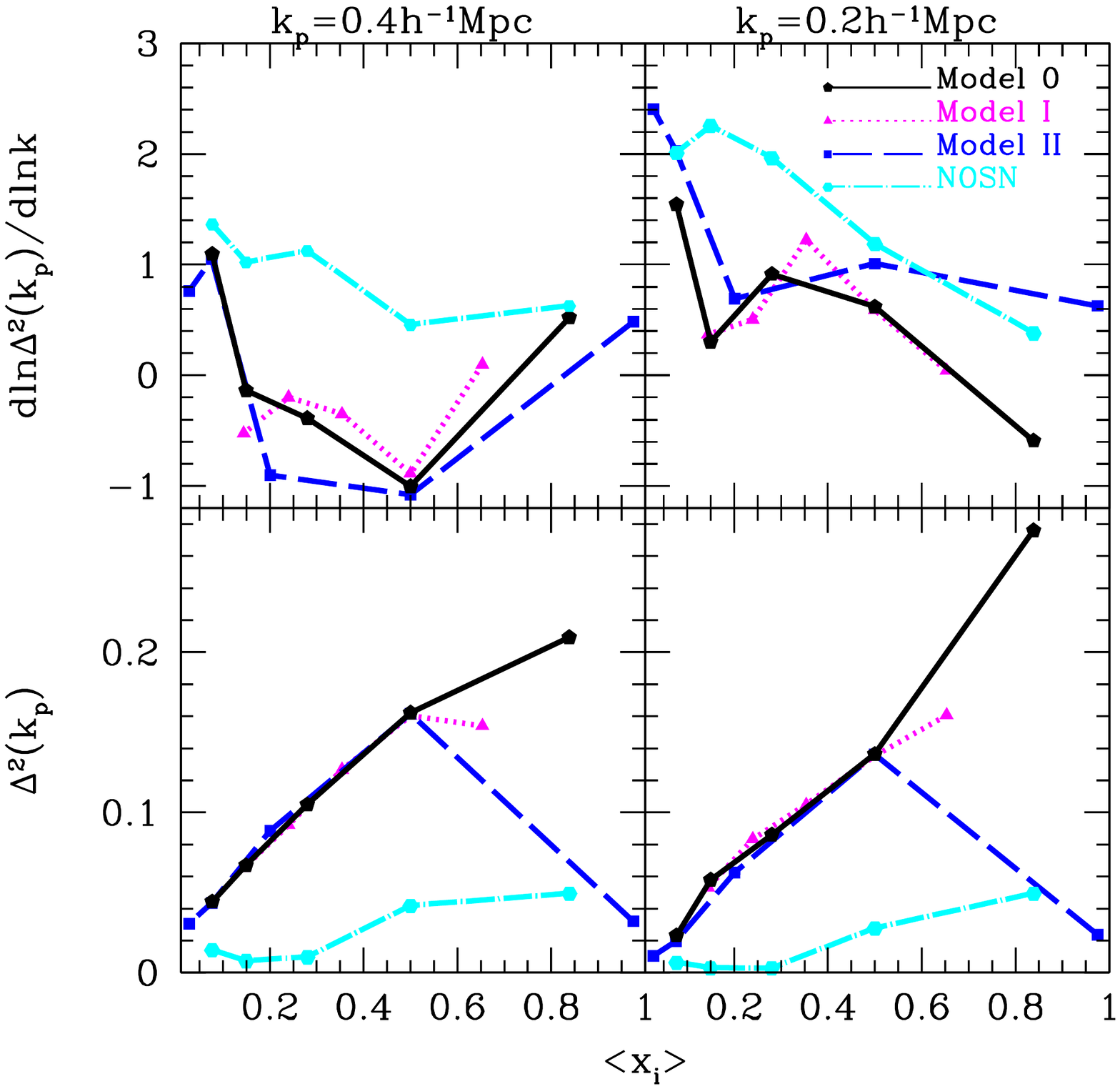}
\includegraphics[width=8cm]{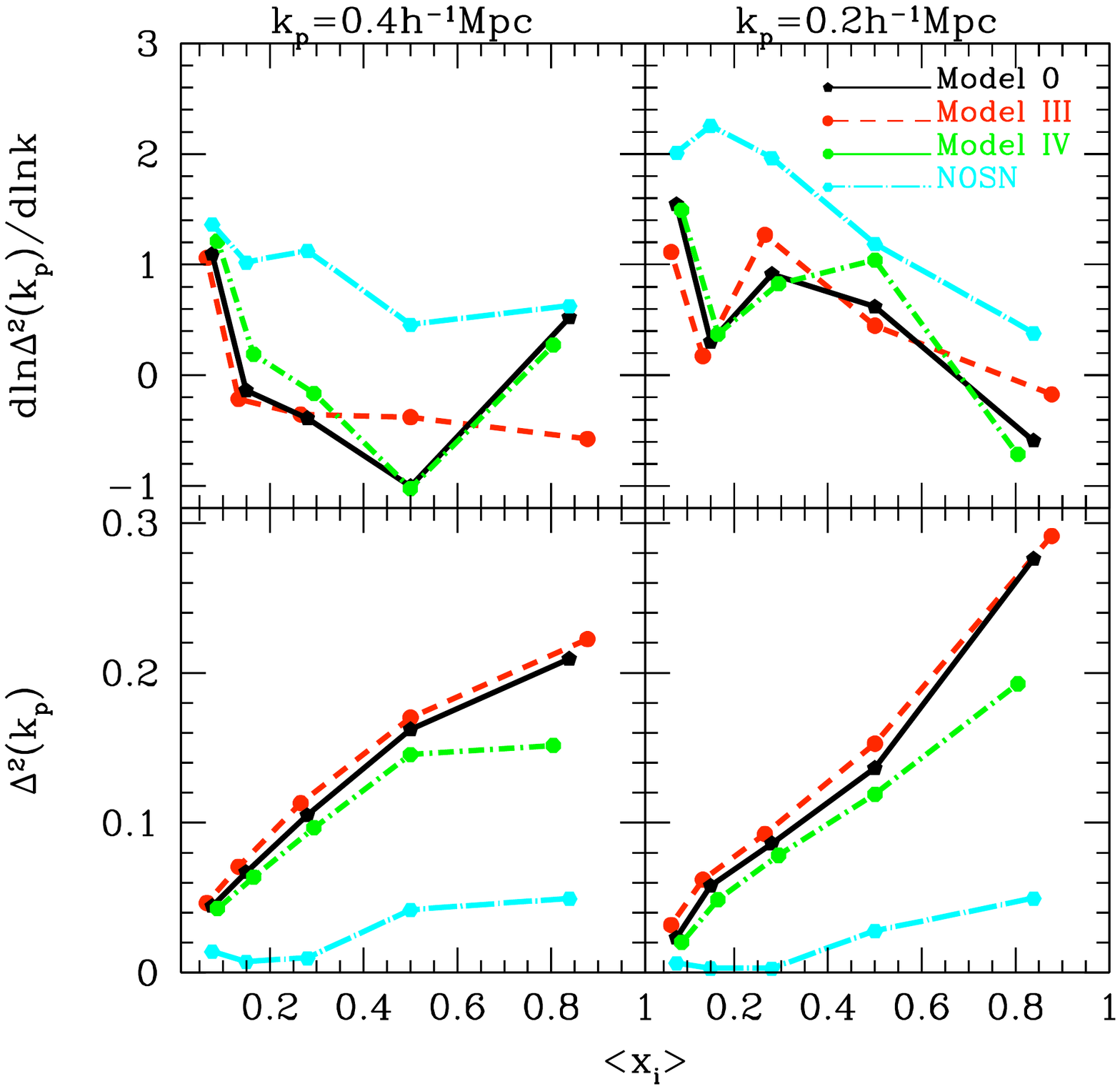}
\end{center}
\vspace{-2.5cm}
\caption{Plots of the evolution in dimensionless 21cm power spectrum amplitude (lower panels) and slope (upper panels) as a function of ionization fraction $\left<x_{i}\right>$. Predictions are shown for  redshift-dependent $f_{\rm esc}$ models (left panel) and  halo mass-dependent $f_{\rm esc}$ models (right panel). Results are shown for two central wave numbers, $k_{\rm p}=0.4h^{-1}$Mpc (left) and $0.2h^{-1}$Mpc (right) in each case, corresponding to the point on the power spectrum at which we evaluate the amplitude and gradient. } \label{COA}
\end{figure*}

\citet[][]{lidz2008} demonstrated that first generation low frequency arrays like the MWA\footnote{www.mwa.org} should have sufficient sensitivity to measure the amplitude and slope of the 21 cm power spectrum. To quantify the effect of assuming an $f_{\rm esc}$ that depends on redshift or host dark matter halo mass we therefore compare the amplitude and slope of predicted 21-cm power spectra for different models. In Figure~\ref{COA},  we plot these values as a function of $\left<x_{i}\right>$ for central wave numbers corresponding to the point on the power spectrum at which we evaluate the amplitude and gradient. We choose results for $k_{\rm p}=0.2h^{-1}$Mpc and $0.4h^{-1}$Mpc, corresponding to the range of wave numbers to be probed by the MWA. We plot both  redshift-dependent (left panel) and host dark matter halo mass-dependent (right panel) models for $f_{\rm esc}$. Interestingly, models 0-IV which describe variation of $f_{\rm esc}$ with host halo mass and redshift do not show significant variation in either amplitude or slope. However the NOSN model shows significant differences across a range of ionization fraction. In particular the amplitude of the NOSN power spectrum is lower than for models 0-IV. 

\subsection{Observational implications}

Since the ionization fraction is not a direct observable, we also plot the progression of a model in the observable plane of power spectrum  amplitude vs slope. These are shown for models 0-IV in Figure~\ref{SLPS}, again for the two values of central wavenumber { $k_{\rm p}$. The arrows show the direction from high to low $\left<x_{i}\right>$. To illustrate the potential for detectability of this difference we also include error bars for wavenumber $k_{\rm p}$=0.4$h^{-1}{\rm Mpc}$ corresponding to estimates for the MWA ({specifically an $r^{-2}$ distribution of 500 antennas}) \citep[][]{lidz2008} assuming 1000 hours integration and 6MHz bandpasses. Figure ~\ref{SLPS} compares models I-IV and NOSN with the Model~0 (which has constant $f_{\rm esc}$). As noted in the previous section, we find that the effect of SNe feedback strength has a much larger influence on the evolution of ionization structure during the EoR than the different dependencies of escape fractions on redshift or host dark matter halo mass considered in this paper. The model with no SNe feedback produces curves in the $\Delta^{2}(k)$-${{\rm dln}\Delta^{2}(k) / {\rm dlnk}}$ plane that lie at lower values of $\Delta^{2}(k)$. Observation of a large $\Delta^{2}(k)$ would therefore imply the presence of SNe feedback (or other mechanism which raises the bias of ionizing sources), irrespective of the dependence on ionization fraction within the range of models considered.
 \begin{figure*}
\vspace{-1.cm}
\begin{center}
\includegraphics[width=8.4cm]{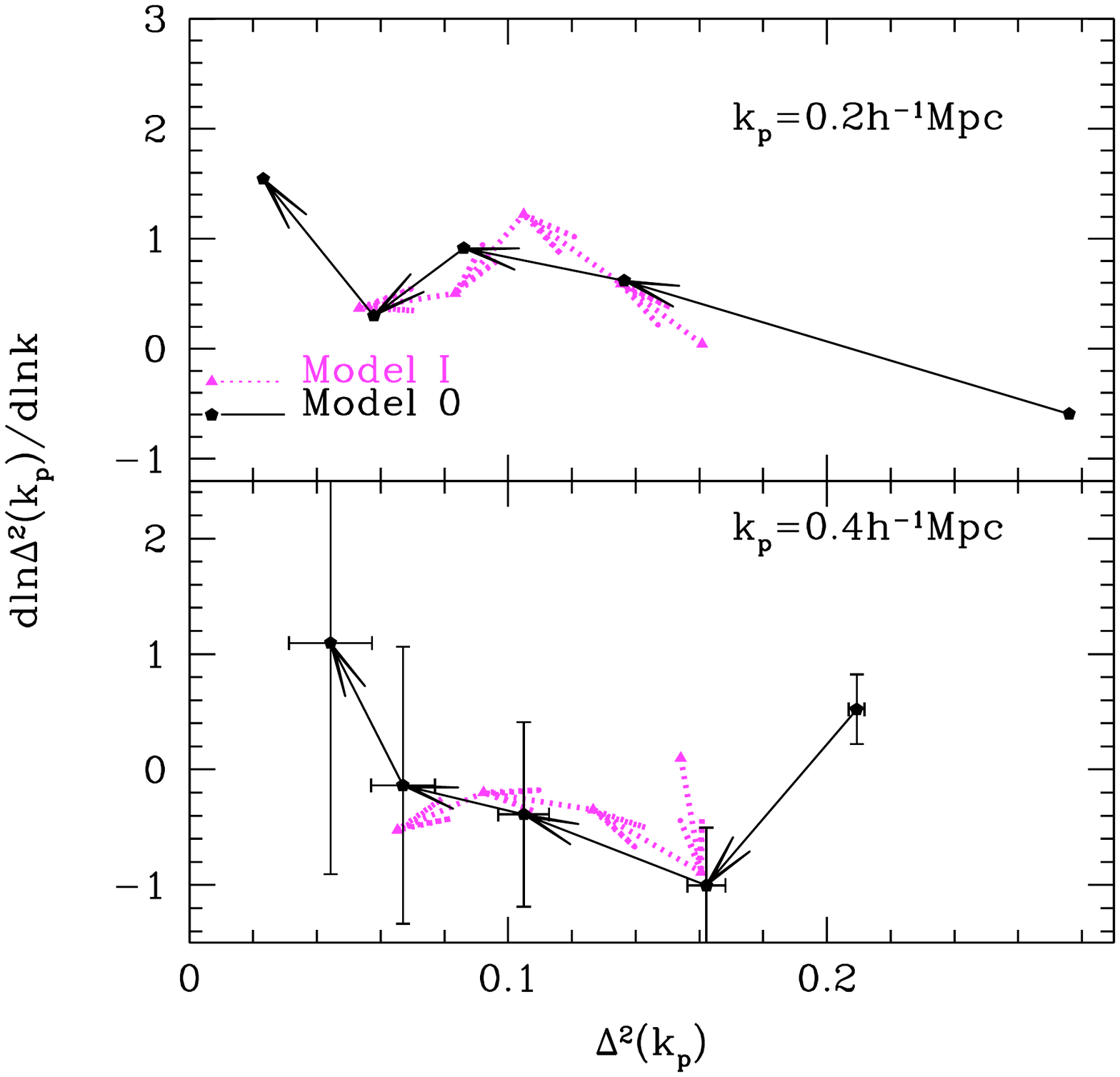}
\vspace{-3.7cm}
\includegraphics[width=8.4cm]{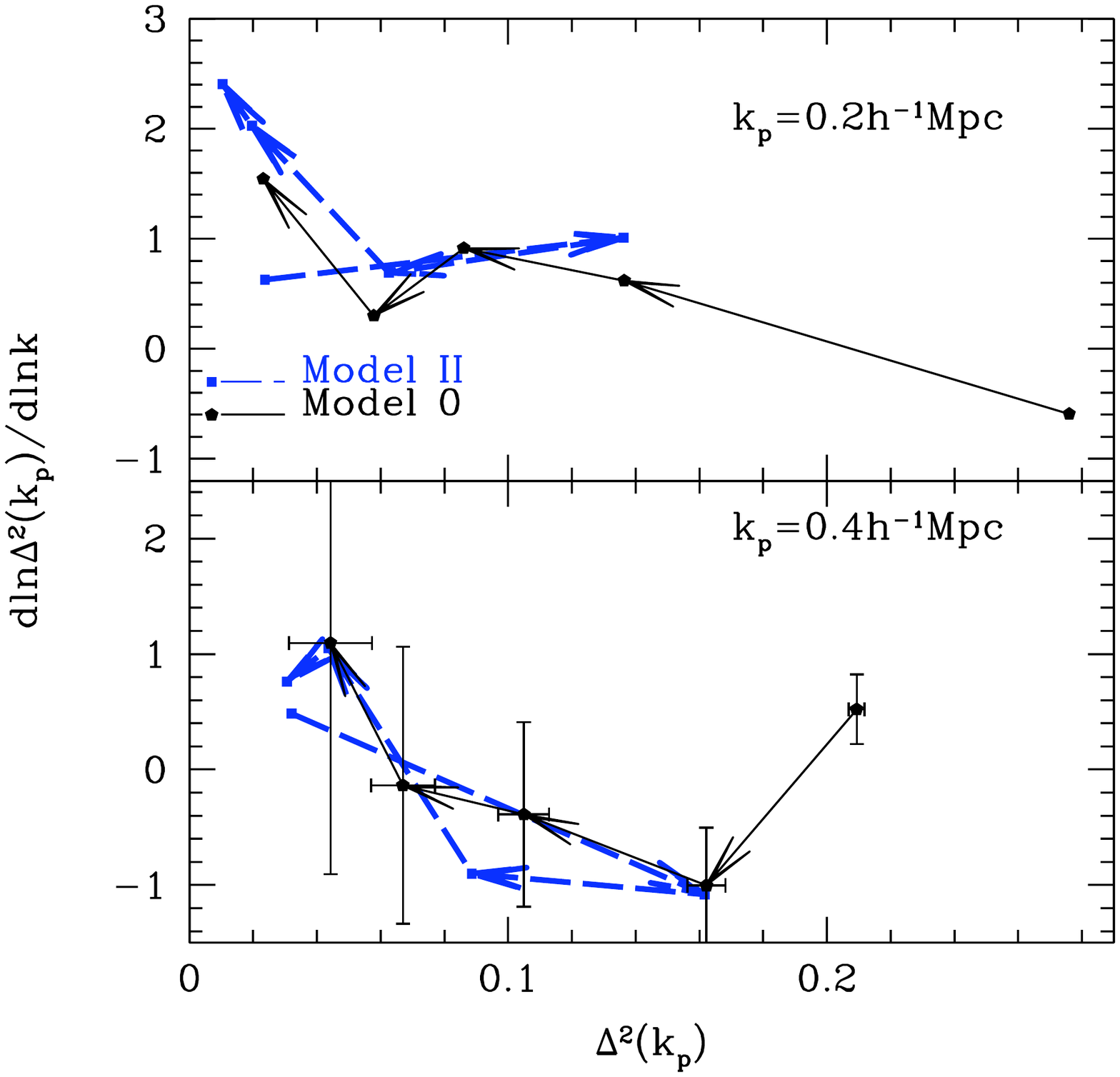}
\includegraphics[width=8.4cm]{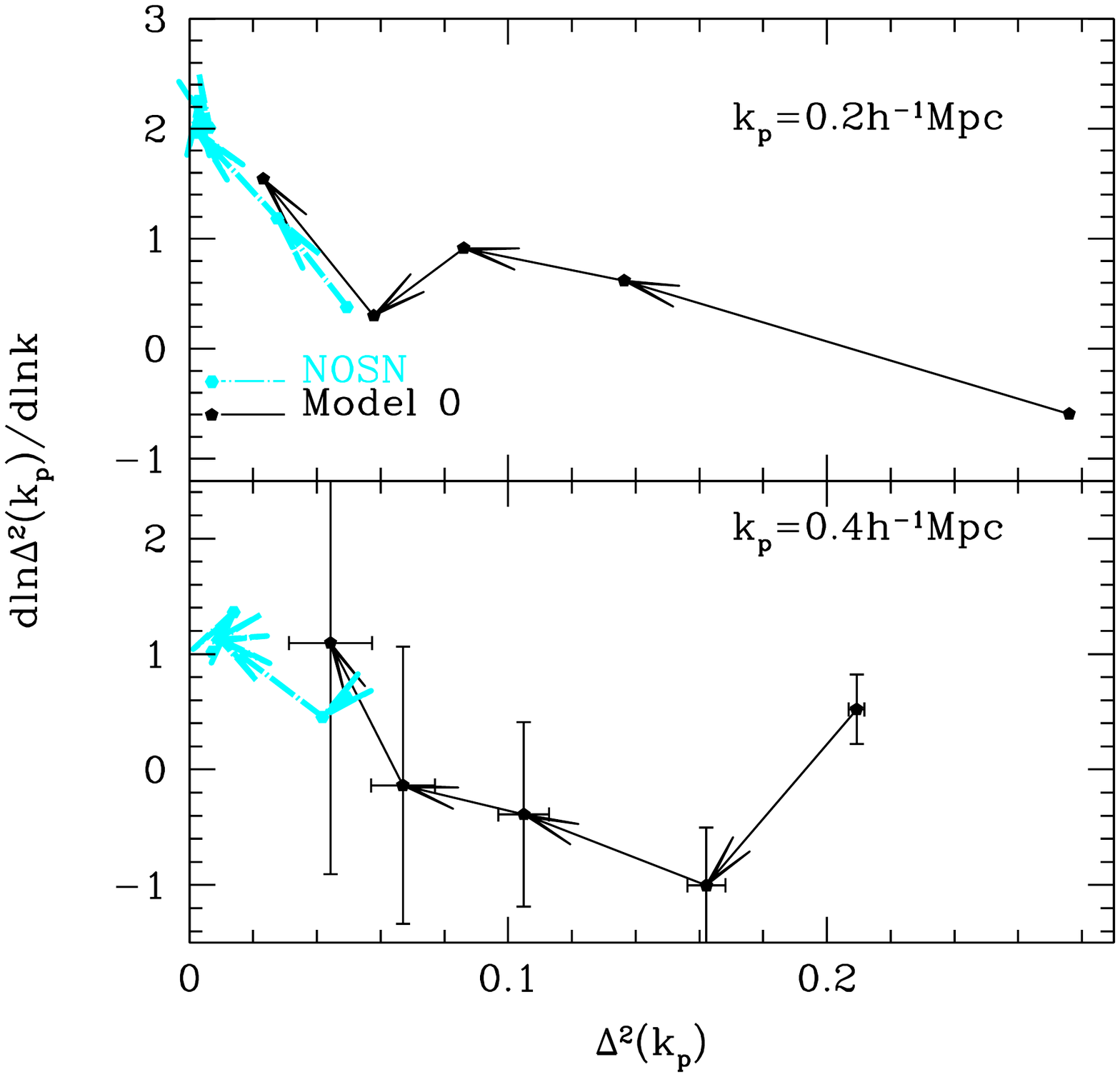}
\vspace{-3.7cm}\\
\includegraphics[width=8.4cm]{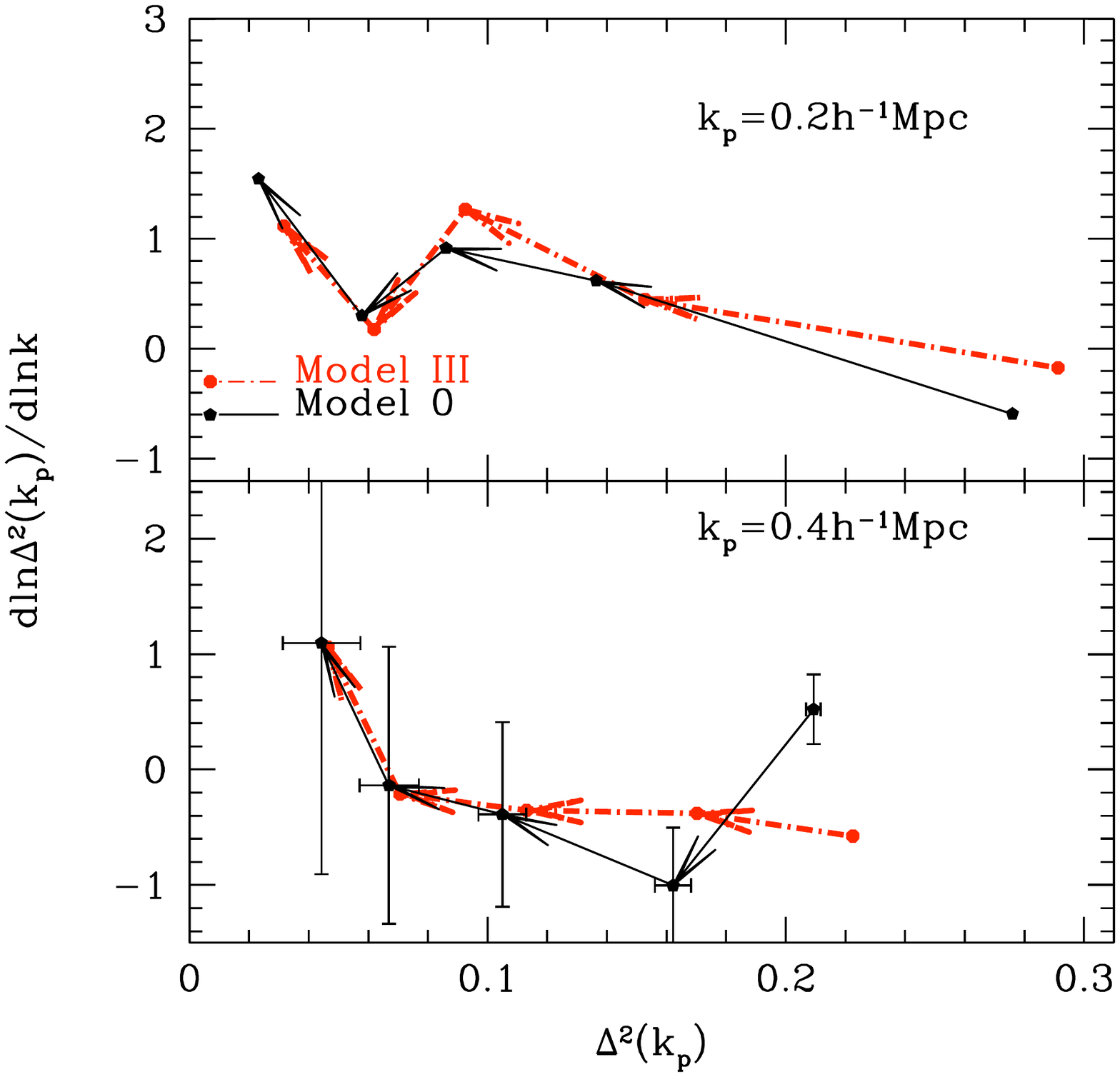}
\includegraphics[width=8.4cm]{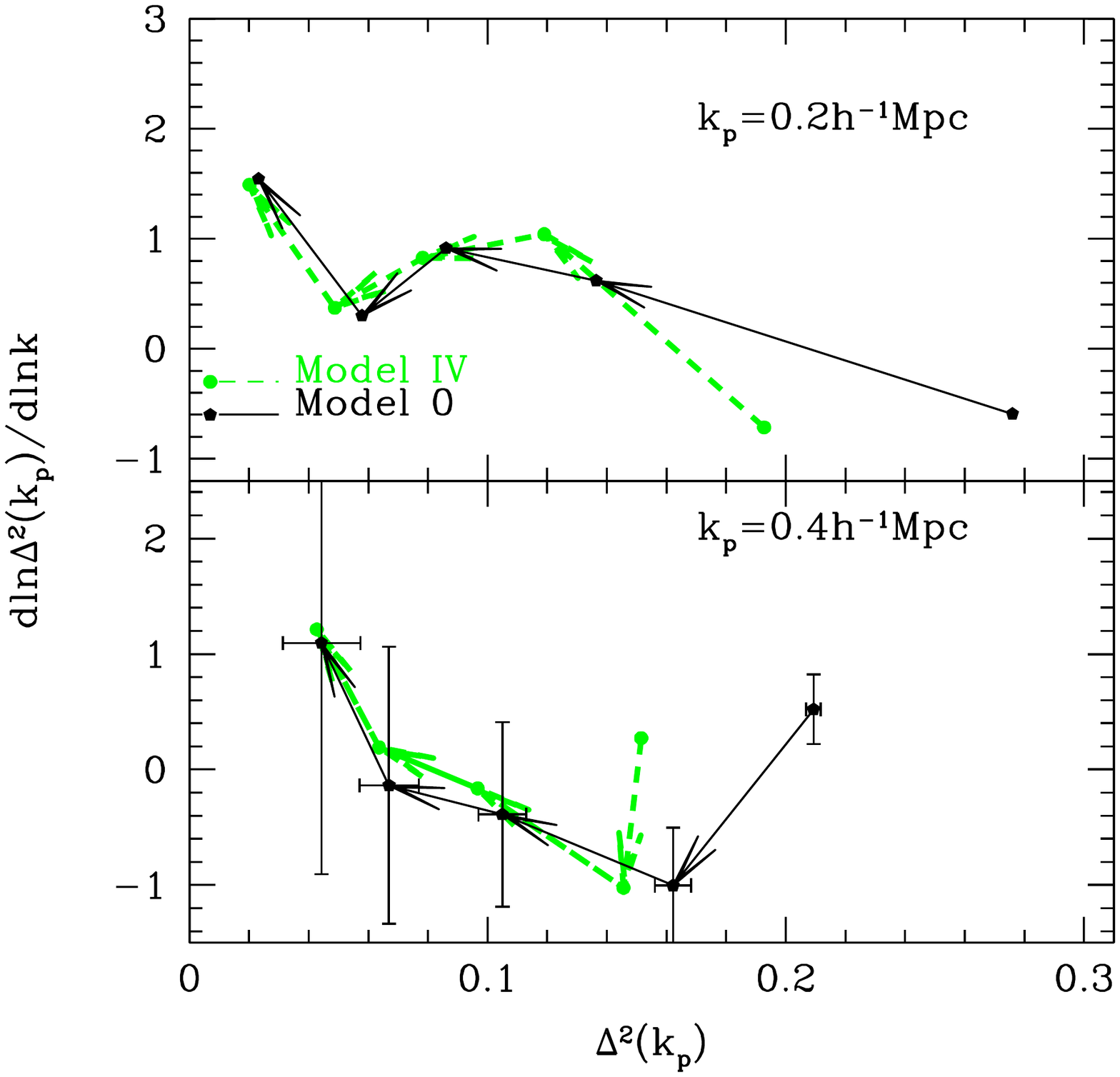}
\end{center}
\vspace{-2.7cm}
\caption{Plots of the loci of points in the parameter space of 21-cm power spectrum amplitude and slope. Loci are shown for each of  redshift-dependent $f_{\rm esc}$ models I and II (top panels), NOSN model (middle panel), and halo mass dependent models III and IV (bottom panels) with the Model~0 prediction plotted for comparison in each panel.
Results are shown for two central wave numbers, $k_{\rm p}=0.2h^{-1}$Mpc (top of each panel) and $0.4h^{-1}$Mpc (bottom of each panel), corresponding to the point on the power spectrum at which we evaluate the amplitude and gradient.
The error bars for $0.4h^{-1}$Mpc at each point on the Model~0 correspond to estimates for the MWA \citep[][]{lidz2008} with 1000 hours of integration and 6MHz of bandpass. }
\label{SLPS}
\end{figure*}

\section{Conclusion}\label{Summary}

Over the next decade we are likely to see the first measurements of the power spectrum of redshifted 21cm fluctuations from neutral hydrogen structure during the Epoch of Reionization. One goal of these experiments will be to learn about the properties of the galaxies that drove the reionization process. It is known that the ionization structure of the IGM, and hence the observed 21cm power spectrum, will be sensitive to the astrophysical properties of the reionizing galaxies. In this paper we have extended the semi-analytic model for reionization presented in \cite{Kim2012a} to include the possibility of redshift or mass dependent escape fractions. This model combines the GALFORM galaxy formation model implemented within the Millennium-II dark matter simulation with a semi-numerical scheme to describe the resulting ionization structure. We find that an escape fraction which varies with galaxy mass and redshift influences the structure of reionization. However, we find that the effect is smaller than the dominant astrophysical influence of SNe feedback. Thus, we conclude that the unknown dependence on the escape fraction will not influence the ability of observations to determine the dominant physics of galaxy formation during reionization from the observed 21cm power spectrum.

\vspace{5mm}

{\bf Acknowledgments} HSK is supported by a Super-Science Fellowship from the Australian Research Council. The Centre for All-sky
Astrophysics is an Australian Research Council Centre of Excellence, funded by grant CE11E0090.  This work was supported in part by the Science and Technology Facilities
Council rolling grant to the ICC. The Millennium-II Simulation was
carried out by the Virgo Consortium at the supercomputer centre of the
Max Planck Society in Garching.
Calculations for this paper were partly performed on the ICC Cosmology Machine, which is part of the DiRAC Facility jointly funded by STFC, the Large Facilities Capital Fund of BIS,
and Durham University.  

\newcommand{\noopsort}[1]{}

\bibliographystyle{mn2e}

\bibliography{21PII-Re}

\begin{thebibliography}{}

\bibitem[\protect\citeauthoryear{{Barkana}}{{Barkana}}{2009}]{barkana2008}
{Barkana} R.,  2009, MNRAS, 397, 1454

\bibitem[\protect\citeauthoryear{{Barkana} \& {Loeb}}{{Barkana} \&
  {Loeb}}{2001}]{BL01}
{Barkana} R.,  {Loeb} A.,  2001, Phys. Rep., 349, 125

\bibitem[\protect\citeauthoryear{{Baugh}}{{Baugh}}{2006}]{Baugh2006}
{Baugh} C.~M.,  2006, Reports on Progress in Physics, 69, 3101

\bibitem[\protect\citeauthoryear{{Baugh}, {Lacey}, {Frenk}, {Granato}, {Silva},
  {Bressan}, {Benson} \& {Cole}}{{Baugh} et~al.}{2005}]{Baugh2005}
{Baugh} C.~M.,  {Lacey} C.~G.,  {Frenk} C.~S.,  {Granato} G.~L.,  {Silva} L.,
  {Bressan} A.,  {Benson} A.~J.,    {Cole} S.,  2005, \mnras, 356, 1191

\bibitem[\protect\citeauthoryear{{Benson}, {Lacey}, {Baugh}, {Cole} \&
  {Frenk}}{{Benson} et~al.}{2002}]{Benson2002}
{Benson} A.~J.,  {Lacey} C.~G.,  {Baugh} C.~M.,  {Cole} S.,    {Frenk} C.~S.,
  2002, \mnras, 333, 156

\bibitem[\protect\citeauthoryear{{Benson}, {Nusser}, {Sugiyama} \&
  {Lacey}}{{Benson} et~al.}{2001}]{Benson2001}
{Benson} A.~J.,  {Nusser} A.,  {Sugiyama} N.,    {Lacey} C.~G.,  2001, \mnras,
  320, 153

\bibitem[\protect\citeauthoryear{{Benson}, {Sugiyama}, {Nusser} \&
  {Lacey}}{{Benson} et~al.}{2006}]{Benson2006}
{Benson} A.~J.,  {Sugiyama} N.,  {Nusser} A.,    {Lacey} C.~G.,  2006, \mnras,
  369, 1055

\bibitem[\protect\citeauthoryear{{Bland-Hawthorn} \&
  {Maloney}}{{Bland-Hawthorn} \& {Maloney}}{1999}]{bland1999}
{Bland-Hawthorn} J.,  {Maloney} P.~R.,  1999, \apjl, 510, L33

\bibitem[\protect\citeauthoryear{{Bower}, {Benson}, {Malbon}, {Helly}, {Frenk},
  {Baugh}, {Cole} \& {Lacey}}{{Bower} et~al.}{2006}]{Bower2006}
{Bower} R.~G.,  {Benson} A.~J.,  {Malbon} R.,  {Helly} J.~C.,  {Frenk} C.~S.,
  {Baugh} C.~M.,  {Cole} S.,    {Lacey} C.~G.,  2006, \mnras, 370, 645

\bibitem[\protect\citeauthoryear{{Chen}, {Prochaska} \& {Gnedin}}{{Chen}
  et~al.}{2007}]{chen2007}
{Chen} H.-W.,  {Prochaska} J.~X.,    {Gnedin} N.~Y.,  2007, ApJL, 667, L125

\bibitem[\protect\citeauthoryear{{Ciardi}, {Stoehr} \& {White}}{{Ciardi}
  et~al.}{2003}]{ciardi2003}
{Ciardi} B.,  {Stoehr} F.,    {White} S.~D.~M.,  2003, MNRAS, 343, 1101

\bibitem[\protect\citeauthoryear{{Cole}, {Lacey}, {Baugh} \& {Frenk}}{{Cole}
  et~al.}{2000}]{Cole2000}
{Cole} S.,  {Lacey} C.~G.,  {Baugh} C.~M.,    {Frenk} C.~S.,  2000, \mnras,
  319, 168

\bibitem[\protect\citeauthoryear{{Fujita}, {Martin}, {Mac Low} \&
  {Abel}}{{Fujita} et~al.}{2003}]{fujita2003}
{Fujita} A.,  {Martin} C.~L.,  {Mac Low} M.-M.,    {Abel} T.,  2003, \apj, 599,
  50

\bibitem[\protect\citeauthoryear{{Furlanetto} \& {Oh}}{{Furlanetto} \&
  {Oh}}{2005}]{furlanetto2006}
{Furlanetto} S.~R.,  {Oh} S.~P.,  2005, MNRAS, 363, 1031

\bibitem[\protect\citeauthoryear{{Gnedin}, {Kravtsov} \& {Chen}}{{Gnedin}
  et~al.}{2008}]{gnedin2007}
{Gnedin} N.~Y.,  {Kravtsov} A.~V.,    {Chen} H.-W.,  2008, ApJ, 672, 765

\bibitem[\protect\citeauthoryear{{Hurwitz}, {Jelinsky} \& {Dixon}}{{Hurwitz}
  et~al.}{1997}]{hurwitz1997}
{Hurwitz} M.,  {Jelinsky} P.,    {Dixon} W.~V.~D.,  1997, \apjl, 481, L31

\bibitem[\protect\citeauthoryear{{Iliev}, {Mellema}, {Pen}, {Bond} \&
  {Shapiro}}{{Iliev} et~al.}{2008}]{Il08}
{Iliev} I.~T.,  {Mellema} G.,  {Pen} U.-L.,  {Bond} J.~R.,    {Shapiro} P.~R.,
  2008, MNRAS, 384, 863

\bibitem[\protect\citeauthoryear{{Iliev}, {Mellema}, {Shapiro} \&
  {Pen}}{{Iliev} et~al.}{2007}]{iliev2007}
{Iliev} I.~T.,  {Mellema} G.,  {Shapiro} P.~R.,    {Pen} U.-L.,  2007, MNRAS,
  376, 534

\bibitem[\protect\citeauthoryear{{Inoue}, {Iwata} \& {Deharveng}}{{Inoue}
  et~al.}{2006}]{inoue2006}
{Inoue} A.~K.,  {Iwata} I.,    {Deharveng} J.-M.,  2006, \mnras, 371, L1

\bibitem[\protect\citeauthoryear{{Kim}, {Baugh}, {Benson}, {Cole}, {Frenk},
  {Lacey}, {Power} \& {Schneider}}{{Kim} et~al.}{2011}]{Kim2011}
{Kim} H.-S.,  {Baugh} C.~M.,  {Benson} A.~J.,  {Cole} S.,  {Frenk} C.~S.,
  {Lacey} C.~G.,  {Power} C.,    {Schneider} M.,  2011, \mnras, 414, 2367

\bibitem[\protect\citeauthoryear{{Kim}, {Power}, {Baugh}, {Wyithe}, {Lacey},
  {Lagos} \& {Frenk}}{{Kim} et~al.}{2013}]{Kim2012b}
{Kim} H.-S.,  {Power} C.,  {Baugh} C.~M.,  {Wyithe} J.~S.~B.,  {Lacey} C.~G.,
  {Lagos} C.~D.~P.,    {Frenk} C.~S.,  2013, \mnras, 428, 3366

\bibitem[\protect\citeauthoryear{{Kim}, {Wyithe}, {Raskutti}, {Lacey} \&
  {Helly}}{{Kim} et~al.}{2013}]{Kim2012a}
{Kim} H.-S.,  {Wyithe} J.~S.~B.,  {Raskutti} S.,  {Lacey} C.~G.,    {Helly}
  J.~C.,  2013, \mnras, 428, 2467

\bibitem[\protect\citeauthoryear{{Kuhlen} \& {Faucher-Gigu{\`e}re}}{{Kuhlen} \&
  {Faucher-Gigu{\`e}re}}{2012}]{kuhlen2012}
{Kuhlen} M.,  {Faucher-Gigu{\`e}re} C.-A.,  2012, \mnras, 423, 862

\bibitem[\protect\citeauthoryear{{Lacey}, {Baugh}, {Frenk} \& {Benson}}{{Lacey}
  et~al.}{2011}]{Lacey2011}
{Lacey} C.~G.,  {Baugh} C.~M.,  {Frenk} C.~S.,    {Benson} A.~J.,  2011,
  \mnras, 412, 1828

\bibitem[\protect\citeauthoryear{{Lagos}, {Bayet}, {Baugh}, {Lacey}, {Bell},
  {Fanidakis} \& {Geach}}{{Lagos} et~al.}{2012}]{Lagos2012}
{Lagos} C.~d.~P.,  {Bayet} E.,  {Baugh} C.~M.,  {Lacey} C.~G.,  {Bell} T.~A.,
  {Fanidakis} N.,    {Geach} J.~E.,  2012, \mnras, 426, 2142

\bibitem[\protect\citeauthoryear{{Lidz}, {Zahn}, {McQuinn}, {Zaldarriaga} \&
  {Hernquist}}{{Lidz} et~al.}{2008}]{lidz2008}
{Lidz} A.,  {Zahn} O.,  {McQuinn} M.,  {Zaldarriaga} M.,    {Hernquist} L.,
  2008, ApJ, 680, 962

\bibitem[\protect\citeauthoryear{{Mellema}, {Iliev}, {Pen} \&
  {Shapiro}}{{Mellema} et~al.}{2006}]{mellema2006}
{Mellema} G.,  {Iliev} I.~T.,  {Pen} U.-L.,    {Shapiro} P.~R.,  2006, MNRAS,
  372, 679

\bibitem[\protect\citeauthoryear{{Mesinger} \& {Furlanetto}}{{Mesinger} \&
  {Furlanetto}}{2007}]{MF07}
{Mesinger} A.,  {Furlanetto} S.,  2007, ApJ, 669, 663

\bibitem[\protect\citeauthoryear{{Putman}, {Bland-Hawthorn}, {Veilleux},
  {Gibson}, {Freeman} \& {Maloney}}{{Putman} et~al.}{2003}]{putman2003}
{Putman} M.~E.,  {Bland-Hawthorn} J.,  {Veilleux} S.,  {Gibson} B.~K.,
  {Freeman} K.~C.,    {Maloney} P.~R.,  2003, \apj, 597, 948

\bibitem[\protect\citeauthoryear{{Rai{\v c}evi{\'c}}, {Theuns} \&
  {Lacey}}{{Rai{\v c}evi{\'c}} et~al.}{2011}]{Theuns2011}
{Rai{\v c}evi{\'c}} M.,  {Theuns} T.,    {Lacey} C.,  2011, \mnras, 410, 775

\bibitem[\protect\citeauthoryear{{Razoumov} \& {Sommer-Larsen}}{{Razoumov} \&
  {Sommer-Larsen}}{2010}]{razoumov2010}
{Razoumov} A.~O.,  {Sommer-Larsen} J.,  2010, \apj, 710, 1239

\bibitem[\protect\citeauthoryear{{Ricotti} \& {Shull}}{{Ricotti} \&
  {Shull}}{2000}]{ricotti2000}
{Ricotti} M.,  {Shull} J.~M.,  2000, \apj, 542, 548

\bibitem[\protect\citeauthoryear{{Santos}, {Amblard}, {Pritchard}, {Trac},
  {Cen} \& {Cooray}}{{Santos} et~al.}{2008}]{S+07}
{Santos} M.~G.,  {Amblard} A.,  {Pritchard} J.,  {Trac} H.,  {Cen} R.,
  {Cooray} A.,  2008, \apj, 689, 1

\bibitem[\protect\citeauthoryear{{Shapley}, {Steidel}, {Pettini}, {Adelberger}
  \& {Erb}}{{Shapley} et~al.}{2006}]{shapley2006}
{Shapley} A.~E.,  {Steidel} C.~C.,  {Pettini} M.,  {Adelberger} K.~L.,    {Erb}
  D.~K.,  2006, \apj, 651, 688

\bibitem[\protect\citeauthoryear{{Shin}, {Trac} \& {Cen}}{{Shin}
  et~al.}{2008}]{shin2008}
{Shin} M.-S.,  {Trac} H.,    {Cen} R.,  2008, ApJ, 681, 756

\bibitem[\protect\citeauthoryear{{Siana}, {Teplitz}, {Colbert}, {Ferguson},
  {Dickinson}, {Brown}, {Conselice}, {de Mello}, {Gardner}, {Giavalisco} \&
  {Menanteau}}{{Siana} et~al.}{2007}]{siana2007}
{Siana} B.,  {Teplitz} H.~I.,  {Colbert} J.,  {Ferguson} H.~C.,  {Dickinson}
  M.,  {Brown} T.~M.,  {Conselice} C.~J.,  {de Mello} D.~F.,  {Gardner} J.~P.,
  {Giavalisco} M.,    {Menanteau} F.,  2007, \apj, 668, 62

\bibitem[\protect\citeauthoryear{{Siana}, {Teplitz}, {Ferguson}, {Brown},
  {Giavalisco}, {Dickinson}, {Chary}, {de Mello}, {Conselice}, {Bridge},
  {Gardner}, {Colbert} \& {Scarlata}}{{Siana} et~al.}{2010}]{siana2010}
{Siana} B.,  {Teplitz} H.~I.,  {Ferguson} H.~C.,  {Brown} T.~M.,  {Giavalisco}
  M.,  {Dickinson} M.,  {Chary} R.-R.,  {de Mello} D.~F.,  {Conselice} C.~J.,
  {Bridge} C.~R.,  {Gardner} J.~P.,  {Colbert} J.~W.,    {Scarlata} C.,  2010,
  \apj, 723, 241

\bibitem[\protect\citeauthoryear{{Sokasian}, {Abel}, {Hernquist} \&
  {Springel}}{{Sokasian} et~al.}{2003}]{sokasian2003}
{Sokasian} A.,  {Abel} T.,  {Hernquist} L.,    {Springel} V.,  2003, MNRAS,
  344, 607

\bibitem[\protect\citeauthoryear{{Trac} \& {Cen}}{{Trac} \&
  {Cen}}{2007}]{trac2007}
{Trac} H.,  {Cen} R.,  2007, ApJ, 671, 1

\bibitem[\protect\citeauthoryear{{Trac}, {Cen} \& {Loeb}}{{Trac}
  et~al.}{2008}]{trac2008}
{Trac} H.,  {Cen} R.,    {Loeb} A.,  2008, ApJL, 689, L81

\bibitem[\protect\citeauthoryear{{Wise} \& {Cen}}{{Wise} \& {Cen}}{2009}]{WC09}
{Wise} J.~H.,  {Cen} R.,  2009, \apj, 693, 984

\bibitem[\protect\citeauthoryear{{Wood} \& {Loeb}}{{Wood} \&
  {Loeb}}{2000}]{wood2000}
{Wood} K.,  {Loeb} A.,  2000, \apj, 545, 86

\bibitem[\protect\citeauthoryear{{Wyithe} \& {Loeb}}{{Wyithe} \&
  {Loeb}}{2007}]{WL07}
{Wyithe} J.~S.~B.,  {Loeb} A.,  2007, MNRAS, 375, 1034

\bibitem[\protect\citeauthoryear{{Yajima}, {Choi} \& {Nagamine}}{{Yajima}
  et~al.}{2011}]{yajima2011}
{Yajima} H.,  {Choi} J.-H.,    {Nagamine} K.,  2011, \mnras, 412, 411

\bibitem[\protect\citeauthoryear{{Zahn}, {Lidz}, {McQuinn}, {Dutta},
  {Hernquist}, {Zaldarriaga} \& {Furlanetto}}{{Zahn} et~al.}{2007}]{zahn2007}
{Zahn} O.,  {Lidz} A.,  {McQuinn} M.,  {Dutta} S.,  {Hernquist} L.,
  {Zaldarriaga} M.,    {Furlanetto} S.~R.,  2007, ApJ, 654, 12

\end{thebibliography}

\label{lastpage}
\end{document}